\documentclass{aa}
\usepackage{graphicx}
\usepackage{epsfig}

\def\kms{\,km\,s$^{-1}$} 
\def\ms{\,m\,s$^{-1}$} 
\def\vsini{$v\sin i$}

\def\ELOD{{\footnotesize ELODIE}}
\def\lxlb{\log(L_{\hbox{\scriptsize X}}/L_{\hbox{\scriptsize bol}})}

\begin{document}

\title{No planet for HD 166435}

\author{D.~Queloz\inst{1}
  \and 
     G.~W.~Henry\inst{2} 
  \and 
     J.~P.~Sivan\inst{3}
  \and 
     S.~L.~Baliunas\inst{2}$^,$\inst{4}$^,$\inst{5} 
  \and 
     J.~L.~Beuzit\inst{6}
  \and 
     R.~A.~Donahue\inst{4}$^,$\inst{5} 
  \and 
     M.~Mayor\inst{1}
  \and 
     D.~Naef\inst{1}
  \and 
     C.~Perrier\inst{6}
  \and 
     S.~Udry\inst{1}}
     
\offprints{D. Queloz}

\institute{Observatoire de Gen\`eve, 51 ch.  des
      Maillettes, CH--1290 Sauverny, Switzerland\\
      \email{Didier.Queloz@obs.unige.ch}
      \and 
      Center of Excellence in Information Systems,
           Tennessee State University,
           330 10th Avenue North,
           Nashville, TN 37203, USA
          \and
           Observatoire de Haute Provence, 
           Saint-Michel l'Observatoire, 04870, France
      \and
           Harvard-Smithsonian Center for Astrophysics,
           60 Garden Street,
           Cambridge, MA 02138, USA
           \and
           Mount Wilson Observatory,
           740 Holladay Road,
           Pasadena, CA 91106 USA
           \and
           Observatoire de Grenoble, 414 rue de la Piscine, 
           Domaine Universitaire de St Martin d'Hi\`eres, 38041, France
      }

\date{Received ; Accepted } 

\abstract{
The G0\,V star HD~166435 has been observed by 
 the fiber-fed spectrograph ELODIE  as one of the targets 
in the large extra-solar planet survey that we are conducting at the 
Observatory of Haute-Provence.  We detected coherent, low-amplitude, 
radial-velocity variations with a period of 3.7987\,days, suggesting a 
possible close-in planetary companion.  Subsequently,  we initiated a series 
of high-precision photometric observations to search for possible planetary 
transits and an additional series of Ca\,II H and K observations to measure 
the level of surface magnetic activity and to look for possible rotational 
modulation.  Surprisingly, we found the star to be photometrically variable 
and magnetically active.  A detailed study of the phase stability of the 
radial-velocity signal revealed that the radial-velocity variability remains 
coherent only for durations of about 30\,days.  Analysis of the time variation 
of the spectroscopic line profiles using line bisectors revealed a correlation 
between radial velocity and line-bisector orientation.  All of these 
observations, along with a one-quarter cycle phase shift between the 
photometric and the radial-velocity variationss, are well explained by the 
presence of dark photospheric spots on HD~166435.  We conclude that the 
radial-velocity variations are not due to gravitational interaction with 
an orbiting planet but, instead, originate from line-profile changes stemming 
from star spots on the surface of the star.  The quasi-coherence of the 
radial-velocity signal over more than two years, which allowed a fair fit 
with a binary model, makes the stability of this star unusual among other 
active stars.  It suggests a stable magnetic field orientation where spots 
are always generated at about the same location on the surface of the star.  
\keywords{stars: activity -- 
         individual: HD166435 -- 
	 planetary systems}
}

\maketitle
%

\section{Introduction}

A large extra-solar planet survey has been underway since 1994 at the 
Observatory of Haute-Provence with the high-precision, fiber-fed echelle 
spectrograph \ELOD~ (\cite{Baranne96}) mounted on the 193\,cm telescope.  The 
search for extra-solar planets is carried out by seeking changes in the 
radial velocity of each star produced by gravitational interaction with 
orbiting planets.  In 1995 this program lead to the first detection of an 
extrasolar planet orbiting a Sun-like star (\cite{MayorQueloz95}).  Our survey 
contains 324 G dwarf stars brighter than $V$ = 7.65, including
HD~166435 (\cite{Quelozetal98b}).

The star HD~166435 is a G0 dwarf with (B$-$V)$=0.633$.  From its distance 
of 25\,pc and its apparent magnitude $m_V=6.85$, we find an absolute 
magnitude $M_V=4.8$.  If the metallicity is roughly solar, HD~166435's 
location in the H-R diagram suggests a main-sequence age of a few billion 
years.  Interestingly, the star was detected in the ROSAT all-sky survey as 
an extreme ultraviolet (EUV) source, but no other evidence of activity was 
found by \cite{Mulliss94} in their analysis of the H$_\alpha$ and 
CaII\,(8542\AA) lines.  \cite{Burleigh97} found a hint of emission in the 
CIV\,(1549\AA) line but detected no other obvious spectral emission features.  
When we selected HD~166435 for ELODIE observations, we had no real indication 
that HD~166435 might be young and active.  A possible coincidence with another 
EUV source had been mentioned (\cite{Burleigh97}) as a possible explanation 
for the EUV ROSAT detection.  

After only a few radial-velocity measurements of HD~166435, it became apparent 
that the star exhibited low-amplitude, radial-velocity variations with a period 
of about 4 days.  Further measurements in the following months confirmed our 
initial finding, leading us to believe that HD~166435 may have a planetary 
companion in a close orbit.  Subsequently, we began a campaign of 
high-precision photometric observations with an automatic photoelectric 
telescope (APT) at Fairborn Observatory in Arizona to search for planetary 
transits and made tentative plans to announce the new planet at the 
Protostars and Planets IV conference (\cite{manetal00}).  To our surprise, 
however, we found the star to be photometrically variable with the same period 
as the radial-velocity variations.  In this article we describe these data, 
along with additional Ca\,II H and K measurements acquired at Mount Wilson 
Observatory, and discuss the most likely explanation for the observed 
variations in all three data sets.

\section{ELODIE spectra}

We obtained 70 spectra with \ELOD\, from 1997 May to 1999 September. These 
spectra have typical signal-to-noise ratios of 70--150 (at 5000\AA) per
resolution element and a resolution ($\lambda/\Delta\lambda$) of about 42,000.  
The data reduction was carried out on-line during the observations by the 
\ELOD~ automatic reduction software (see \cite{Baranne96} for details).  All 
the spectra were acquired in the high-precision mode, which provides a 
simultaneous thorium reference spectrum.  The zero-point wavelength 
calibration has an intrinsic instrumental precision of 10\ms.

To examine certain key spectral features, we co-added the 70 individual 
spectra to obtain a composite spectrum with a very high signal-to-noise ratio 
of about 1000.  Since the individual spectra were sampled at slightly 
different velocities relative to the star due to the earth's orbital motion, 
the composite spectrum also has a  better quality than the 
individual spectra.  Two key spectral features in the composite spectrum are 
shown in Fig.\,1.  The Ca\,II H spectral region shows an emission reversal in 
the core of the absorption line, suggesting strong chromospheric activity.  
The broader Ca\,II photospheric wings compared to the Sun are the consequence 
of the higher effective temperature of HD~166435.  No strong lithium 
absorption line is seen.  The comparison with the Sun's spectrum suggests 
that, with our resolution, the sensitivity to the $^7$Li doublet is limited to 
medium or strong features.  Moreover, in the case of HD\,166435, the Lithium 
doublet is blended with a nearby feature due to rotational broadening.  
However, it appears that the $^7$Li feature in HD~166435 is no
stronger than the one in the solar spectrum.  We estimate an upper limit of 
10\,m\AA~ for the $^7$Li doublet at 6707.8\AA~.  The curve of growth from 
\cite{Sod93a} then implies an upper limit for the lithium abundance of 
N(Li)$<1.7$. 

\begin{figure}
\psfig{width=\hsize,file=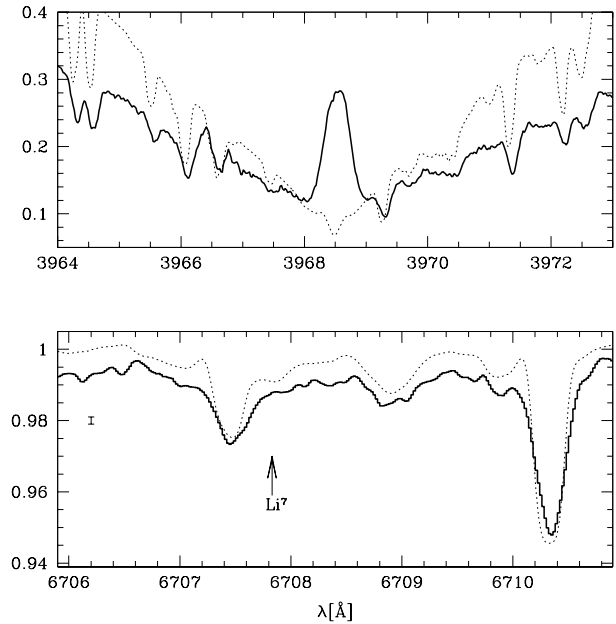}
\caption[]{Selected regions of the composite spectra of HD~166435 (solid 
line).  For comparison, a solar  integrated spectrum (\cite{Kurucz84}) 
 with the same resolution has been 
superimposed (hatched line). {\bf Upper panel} Ca\,II H line.  An emission 
feature is clearly visible in the core of the line. {\bf Lower panel} Region 
of the Li$^7$ line at 6708\AA~. The solar spectrum is slightly offset upwards 
for display purpose.  No strong Li$^7$ feature is detected in the HD\,166435 
spectrum. Notice in both figures the larger line broadening 
of HD 166435 compared to the Sun}
\end{figure}
 
We computed the radial velocity of each individual spectrum with the \ELOD~ 
automatic reduction software, which uses a cross-correlation technique with 
binary mask templates.  In addition to the stellar radial velocity, the 
cross-correlation function (CCF) provides additional information about 
features in the averaged spectral lines (see \cite{Queloz94} for a review).  
The width of the CCF yields the \vsini~ of the star, while the equivalent width 
of the CCF can be used as a metallicity estimate if the temperature of the 
star is known approximately (\cite{Mayor80}; \cite{Benz81}; \cite{Queloz94}). 
The mean equivalent width of the CCF for the 70 individual spectra is similar 
to the value measured for the Pleiades stars of the same temperature 
(B$-$V$_0=0.63$).  Therefore, a solar metallicity can be assumed for 
HD~166435.  With the same technique and calibration described in 
\cite{Quelozetal98a}, we dervive a mean \vsini$=7.6\pm0.5$\kms~. A similar 
value is measured by CORAVEL (\vsini$=7.7\pm0.7$\kms).  

%
%

\subsection{Radial-velocity data}

Analysis of our entire two-year radial-velocity data set revealed a 
periodicity of 3.7987($\pm.0004$) days.  The radial velocities are shown 
in Fig.\,2 where they are phase folded on that period.  We fit these data 
with a binary model and derived a slight orbital eccentricy $e=0.2$ and an 
amplitude $K=83$\ms, corresponding to a minimum mass for a planetary companion 
of 0.6\,$M_J$.  The rms of the residuals to the fit is 28\ms. A large value
compared to the typical \ELOD~ precision of 10\ms.  If the binary model is the correct 
interpretation of the observed radial-velocity variations, then additional 
radial-velocity scatter intrinsic to the atmosphere of the star must 
be present.

\begin{figure}
\psfig{width=\hsize,file=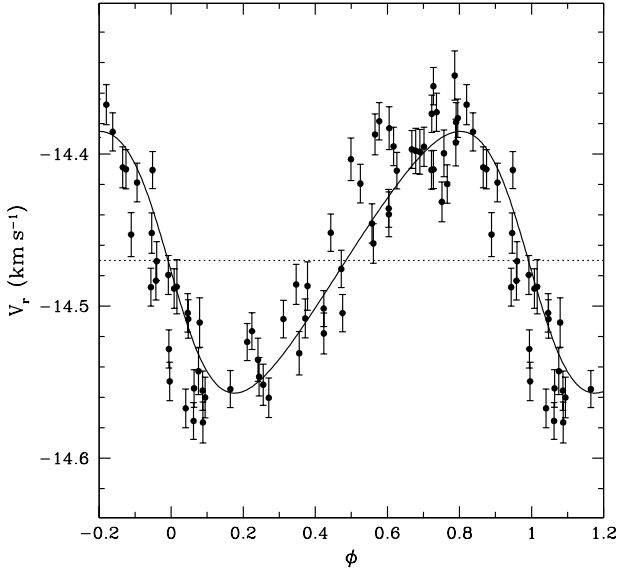}
\caption[]{Phase diagram of the radial-velocity data with a period of 
3.7987\,day.  A binary model is superimposed on the data (solid line). 
See text for details.} 
\end{figure}

In order to test the binary hypothesis, we first investigated the 
coherence time of the radial-velocity variations.  Since a short-period
planet in a (presumed) circular orbit  induces sinusoidal reflex motion  
in the star, we fit sine curves with a period of 3.7987\,days to subsets of 
the radial-velocity observations of increasing duration and computed the rms 
of the residuals in each case.  The results of this test are shown in Fig.\,3. 
The rms of the sinusoidal fits increase with an increase of the duration 
of the data sets.  For durations less than 30\,days, the residuals are 
in agreement  with the expected rms by the photon and instrumental noise.  
For durations longer than 30\,days, 
phase disruptions likely occur and limit the phase coherence and therefore 
increase the rms of the fit.
  
\begin{figure}
\psfig{height=\hsize,angle=270,file=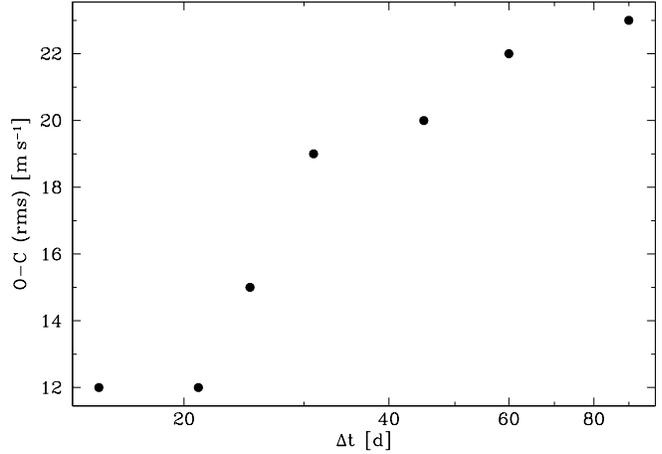}
\caption[]{Variation of the rms (O-C) of a sine-curve fit to the radial-velocity
data versus the time duration of the observations.}
\end{figure}

To quantify the phase disruptions in the radial-velocity data, we searched 
for phase changes among 21-day blocks of data spread over the two-year 
duration of our observations.  From Fig.\,3 above, we see that, for data 
sets of three weeks duration, the radial-velocity variations are coherent 
enough to be described as a sinusoid to within the precision of the 
measurements.  The top panel of Fig.\,4 shows phase shifts of up to 
$\pm~0.1$ cycle among the 21-day blocks of data.  For comparison, we applied 
the same procedure to the 51\,Peg radial-velocity data, where the circular 
orbital motion of this confirmed planet in close orbit produces sinusoidal 
reflex motion.  As expected, no significant phase changes can be seen in 
the 51\,Peg data (Fig.\,4, bottom panel).  Interestingly, the random phase 
changes in the radial velocities from HD\,166435, while signifcantly larger 
than in 51\,Peg, are not so large as to prevent their reasonably good 
description by a sinuosidal model of the whole two-year data set.

\begin{figure}
\psfig{width=\hsize,file=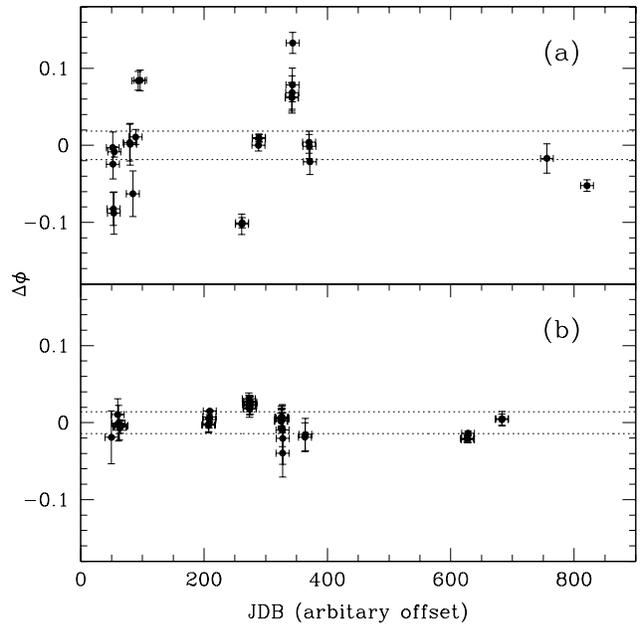}
\caption[]{Phase shifts in the radial velocities derived by fitting sine 
curves to subsets of the data with a maximum duration of 21 days. {\bf (a)} 
HD~166435 data.  {\bf (b)} 51\,Peg data.  Note that there is an arbitrary 
time offset between the two data sets for display purposes.  The two hatched 
lines in each panel indicate the $\pm1\sigma$ error on the global fit.} 
\end{figure}

The detection of significant phase shifts in the HD~166435 radial velocities 
explains the high rms (28\ms) of the residuals to the planet model  fit in
Fig.\,2.  We also examined the amplitude $K$ of the  fits and their 
zero levels $V_0$ for similar changes with the same technique.  No significant 
change of the $K$ amplitude was detected for either star.  The $V_0$ 
values for HD~166435 exhibited small variations that are probably related to 
line-profile variations (see below).  

\subsection{Line profile and bisector analysis}

The radial velocity of a star is defined to be the velocity of the center of 
mass of the star along our line of sight.  However, the observational 
determination of a star's radial velocity is accomplished by measuring the 
Doppler shift of spectral lines produced in the stellar atmosphere.  If
changes occur in the star's spectral-line profiles, the measured Doppler 
shifts may not correspond precisely to the velocity of the star's 
center of mass.  For pulsating stars and stars with very active atmospheres, 
this is indeed the case.  

To interpret observed radial-velocity variations as true changes in the 
velocity of the star, we must show that the observed variations do not 
stem from changes in the stellar atmosphere.  One of the best ways to do
this is to look for subtle changes in the line profiles.  Radial-velocity
changes in a star accompanied by spectral-line profiles that are constant in 
time would be a direct evidence for real velocity changes in the star's
center of mass.  The detection of line-profile changes would point instead 
to atmospheric phenomenon.

The computation of a stellar radial velocity from a particular spectrum with 
the CCF technique involves the correlation of thousands of spectral lines.
Thus, in order for line-profile changes to affect the radial-velocity
measurements, the profile changes in a particular spectrum must affect
all of the spectral lines in a similar way.  Otherwise, no change would
be induced in the cross-correlation function. 
  
Our CCF technique employs a binary template (\cite{Queloz94}) and is a 
very efficient way to look at global phenomena in spectra.  The CCF function 
represents a mean spectral-line profile of all of the lines selected by
the template.  To study the mean spectral-line profile of HD~166435, we 
designed a specific template that selects only the weak and non-saturated 
spectral lines.  We used the line-bisector technique to analyze 
the CCF profiles in the same way that we would analyze an individual spectral 
line.  In Fig.\,5 we show HD~166435's mean CCF and its bisector.  The bisector 
displays the classic ``C'' shape observed in other stars lying on the same 
(cool) side of the granulation boundary (\cite{Gray89}).  Unlike similar 
studies (\cite{Toner88}), our data have an absolute wavelength calibration 
with an accuracy of approximately 10\ms~ (the precision of the instrument).  
Therefore, we don't need to shift the bisector arbitrarily to a mean value 
for comparison.  Moreover, if a bisector change is detected, we know whether 
it is the top or the bottom of the bisector that is moving.  

\begin{figure}
\psfig{width=\hsize,file=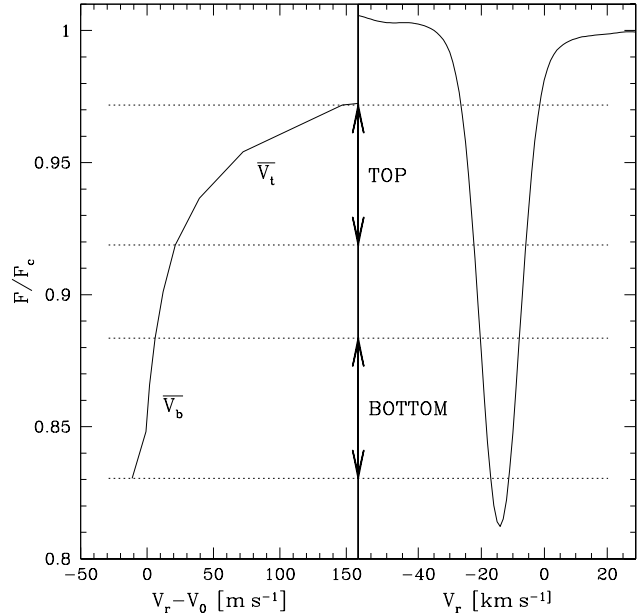}
\caption[]{{\bf Right} The mean CCF function of HD~166435's spectra constructed 
with a template selecting only the weak and non-saturated lines.  This profile 
represents the mean spectral-line profile of the lines selected by the 
template. {\bf Left} The bisector of the CCF.  $V_0$ is an arbitrary offset. 
Note the definition of the boundaries for the computation of 
($\overline{V_t}$ and $\overline{V_b}$).} 
\end{figure} 

We selected two regions of the bisector at the top and bottom of the profile 
($\overline{V_t}$ and $\overline{V_b}$) (see Fig.\,5 for illustration) to 
measure possible bisector orientation changes with high accuracy.  The 
difference between $\overline{V_t}$ and $\overline{V_b}$ is equivalent to 
the "bisector velocity span" widely used in similar studies (see 
\cite{Toner88}).  This is actually a measurement of the inverse of the mean 
slope of the bisector. 
 
Any correlation between radial-velocity changes and line-bisector orientation 
would cast serious doubts on the reflex-motion interpretation of the 
radial-velocity variations.  If the radial-velocity variations are due to 
center-of-mass velocity changes of the star, we would see the bisectors 
oscillating to the left and right of the mean bisector with no change in 
shape or orientation.  In order to investigate this, we first looked at the 
bisectors of each individual CCF.  In Fig.\,6 the bisectors of the CCFs for 
two sets of spectra selected at opposite phases of the radial-velocity cycle 
($\phi=0.0\pm0.1$ and $\phi=0.5\pm0.1$) are displayed.  Comparison with the 
mean bisector shows that the bisectors are twisting around the mean bisector 
with the same periodicity as the radial-velocity signal.  The top of the 
CCF profile moves very little, while the bottom shifts back and forth with 
an amplitude 50\% larger than the $K$ amplitude of the radial-velocity 
signal.  The only way to understand this twisting motion of the bisectors is 
through a change in the spectral-line profiles synchronized with the 
radial-velocity cycle.  The amplitude of this effect is enough to produce 
the observed radial-velocity signal.
 
\begin{figure}
\psfig{width=\hsize,file=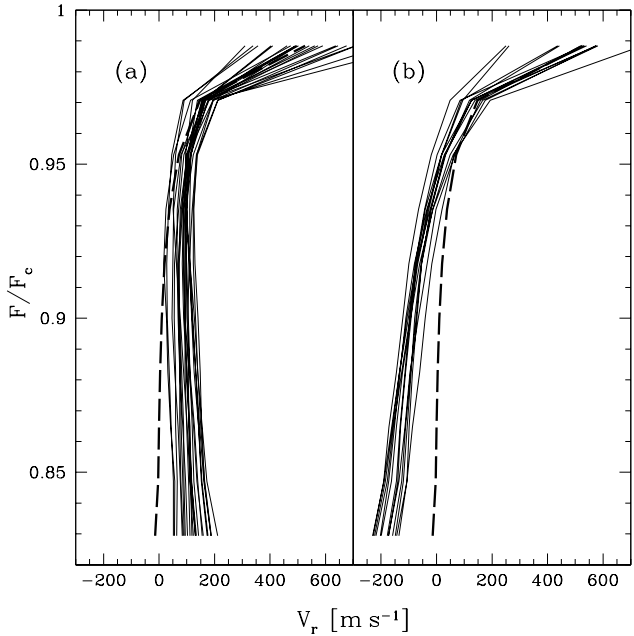}
\caption[]{Individual bisectors for two sets of spectra selected at opposite 
phases of the radial-velocity cycle. {\bf a} Spectra measured at 
$\phi=0.0\pm0.1$. {\bf b} Spectra measured at $\phi=0.5\pm0.1$. The hatched 
line illustrates the mean bisector computed by averaging all spectra.} 
\end{figure}

To investigate further the exact relationship between the orientation 
of the bisectors (bisector span) and the radial velocity of the CCF, 
we plot these two parameters in Fig.\,7.  A direct relationship between 
the two quantities is clearly visible.  A linear solution 
$(\overline{V_t}-\overline{V_b})=-0.88(\pm0.04) V_r$ can be fit with 
a 27\ms~ rms.  Thus, the results of our line-profile analysis of HD~166435's
CCFs provide strong evidence that the radial-velocity variations originate
in the stellar atmosphere and not from reflex motion of the whole star.

\begin{figure}
\psfig{width=\hsize,file=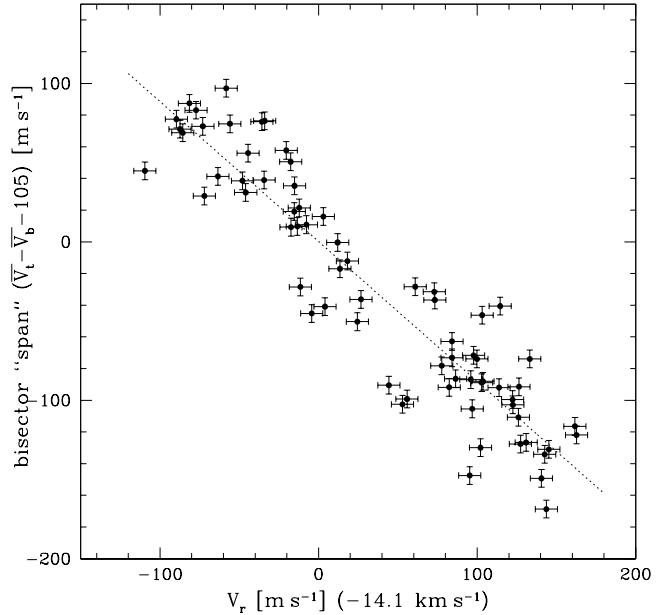}
\caption[]{Radial velocity of each CCF versus the bisector span 
$(\overline{V_t}-\overline{V_b})$ of the CCF profile. The dotted line is 
the best linear fit to the data.} 
\end{figure} 

\section{Photometry}

The {\it Hipparcos} catalogue lists 144 photometric measurements of HD\,166435 
acquired between 1990 January and 1993 March and identifies the star as an 
``unsolved'' variable with a range of 0.05~mag (\cite{Perrymanetal97}).  Our 
periodogram analysis of the {\it Hipparcos} photometry confirms the lack of 
periodicity, particularly near the radial-velocity period.  This is what led 
us initially to suspect that our observed radial-velocity variations might be 
due to the presence of a planet in an orbit similar to that of the companion 
of 51~Pegasi (\cite{MayorQueloz95}). 
 
Thus, shortly after the detection of the periodicity in the radial-velocity 
signal, we began photometric observations of HD\,166435 with the T8 0.80\,m 
automatic photoelectric telescope (APT) at Fairborn Observatory in Arizona 
to search for possible transits of the companion across the disk of the star.  
We acquired 326 Str\"{o}mgren $b$ and $y$ observations with a two-channel 
precision photometer on the APT between 1998 June and 2000 June.  The 
observations were reduced differentially with respect to the comparison star 
99 Her (HR\,6775, HD\,165908, F7V), corrected for atmospheric extinction with 
nightly extinction coefficients, and transformed to the Str\"{o}mgren system.  
External precision of a single observation with the 0.80~m APT averages 
0.0011~mag.  Further details of the automatic-telescope operations and 
data-reduction procedures can be found in \cite{Henry99}.  The individual 
photometric observations are available on the Tennessee State University 
Automated Astronomy Group web site\footnote{ 
http://schwab.tsuniv.edu/t8/hd166435/hd166435.html}.
  
Periodogram analysis of the entire set of 326 observations taken together 
reveals a photometric period of $3.7995 \pm 0.0005$\,d.  Thus, the 
photometric and radial-velocity periods agree within their respective
uncertainties.  

For our photometric analysis we adopt, as reference, the radial-velocity 
period of 3.7987-day and $T_0=2450996.5$, the radial-velocity maximum of 
the best-fit sine-curve model.  The photometric observations are divided into 
five groups as shown in Table~1, where we give the results of least-squares, 
sine-curve fits on the radial-velocity period to the five Str\"{o}mgren $y$ 
data sets.  We also list the periods derived from the five individual data 
sets.  The photometric amplitudes vary from 0.050\,mag to 0.035\,mag; the 
mean magnitudes of the five data sets have a range of about 0.023\,mag.  
Although the five light curves approximate sinusoids, the rms values are, 
nevertheless, somewhat higher than the 0.0011\,mag precision of typical 
observations.  This is primarily the effect of slight cycle-to-cycle changes 
in the light curves within each data set; the rms values increase for the 
later data sets since the later data sets are longer and contain more cycles.
  
\begin{table*}[]
\caption[]{Summary of the Str\"{o}mgren $y$ Photometric Results}
\begin{tabular}{cccccccc}
\hline
\noalign{\smallskip}
Data & Date Range & N$_{\rm obs}$ & Period & Mean Brightness & Full Amplitude & Phase of & rms \\
Set & (JD $-$ 2400000) & & (days) & (mag) & (mag) & Minimum & (mag) \\
\noalign{\smallskip}
\hline
1 & 50986--50996 & 219 &  3.732(6) & 1.849(1) & 0.050(1) & 0.20(1) & 0.003 \\
2 & 51086--51136 &  16 &  3.776(7) & 1.865(1) & 0.052(4) & 0.12(1) & 0.004 \\
3 & 51224--51360 &  42 & 3.807(10) & 1.857(1) & 0.025(4) & 0.07(3) & 0.008 \\
4 & 51434--51501 &  25 & 3.763(34) & 1.842(2) & 0.023(4) & 0.23(3) & 0.007 \\
5 & 51591--51710 &  24 &  3.821(7) & 1.842(2) & 0.032(6) & 0.29(3) & 0.008 \\
\noalign{\smallskip}
\hline
\end{tabular}
\end{table*}

Data set 1, which covers only 10 days or 2.7 cycles, has the largest number 
of observations since at this time we were making repeated measurements 
each night to search for transits.  These data are plotted in the left and 
right third panels from the top of Fig.\,8.  Based on the {\it Hipparcos} 
photometry mentioned above, we did not expect to find coherent light 
variability on this period.  Instead, we found a smoothly varying, nearly 
sinusoidal light curve with a period matching the radial-velocity period 
and a minimum near phase 0.25, where we had hoped to find transits.  
 
\begin{figure*}
\psfig{height=\hsize,angle=-90,file=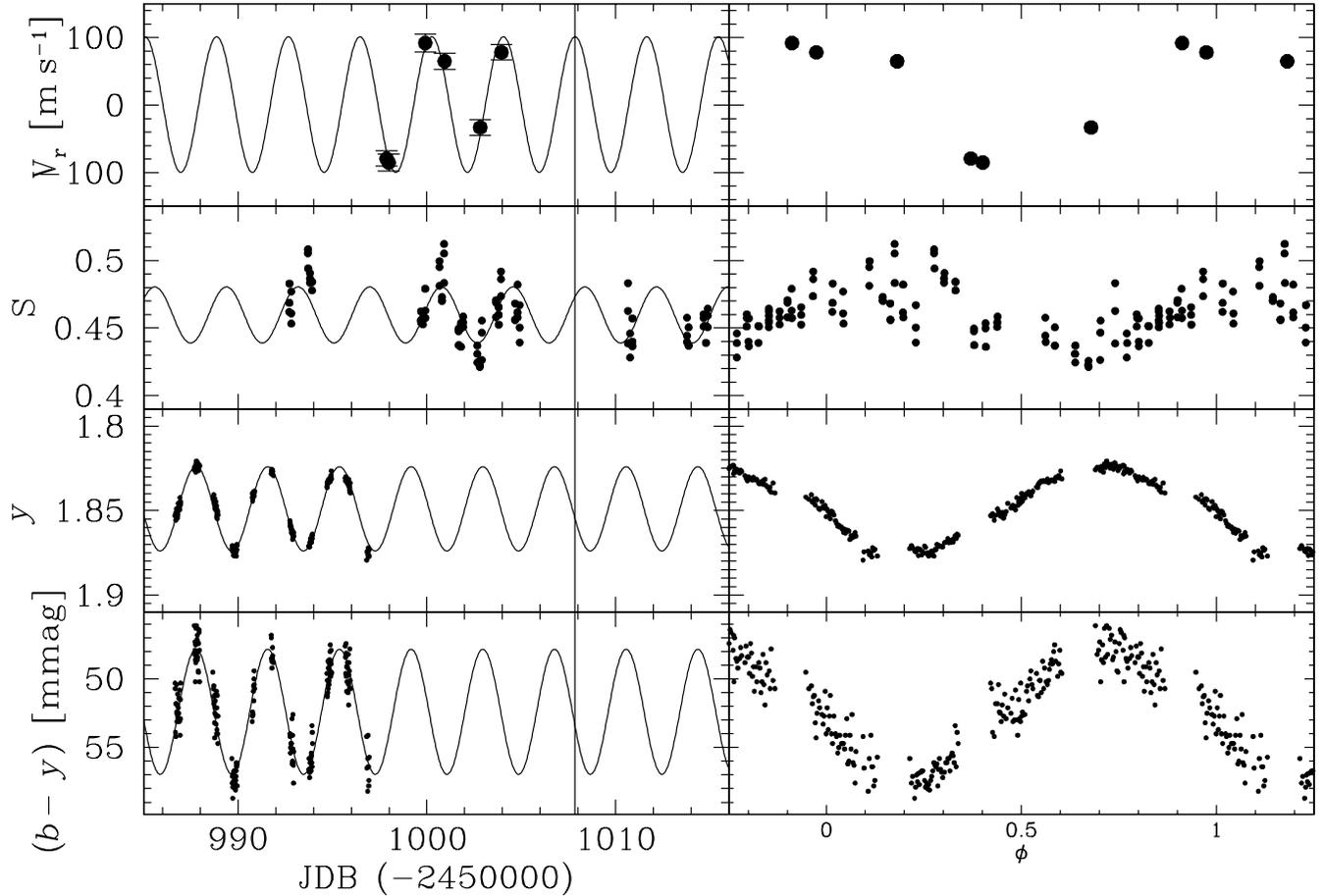}
\caption[]{{\bf Left} Simultaneous observations of (from the top) radial 
velocity, S index, delta $y$ magnitude, and delta $(b-y)$ color, of HD~166435 
over a time span of 30 days.  A best-fit sine-curve with a period fixed at  
3.798 days is shown.  A vertical line is drawn at an epoch of maximum 
radial velocity to help visualize the phase offsets between data sets. 
{\bf Right} Same data but phase folded with $P=3.798$\,d and $T_0=2450996.5$.} 
\end{figure*} 

In Fig.\,8 we see that the time of radial-velocity maximum, i.e., maximum 
red shift at phase 0.0, corresponds closely to the time when the light curve 
crosses its mean level at $\Delta y = 1.85$\,mag in the downward direction.  
The time of photometric minimum corresponds closely to the time when the 
radial-velocity curve crosses its mean level (zero velocity shift with 
respect to the mean velocity) in the downward direction.  The time of 
radial-velocity minimum corresponds closely to the time when the light 
curve crosses its mean level in an upward direction.  Finally, the 
light-curve maximum corresponds closely to the time when the radial-velocity 
curve crosses its mean level in an upward direction.  Hence, these two 
nearly sinusoidal curves are one-quarter cycle out of phase with each other.
  
Our Str\"{o}mgren photometry and $\Delta (b-y)$ colors from data set 1 are 
plotted in the bottom two panels of Fig.~8, also with phases computed from the 
radial-velocity ephemeris.  The color curve is exactly in phase with the 
light curve and has a peak-to-peak amplitude of 0.009\,mag.  The star is 
redder when it is fainter.

\section{Ca II H and K Spectrophotometry}

From 1998 June to 1998 August, 141 Ca II H and K measurements were made on
20 nights with the 100-inch telescope at Mount Wilson Observatory as part of 
the HK Project (\cite{Baliunasetal98}).  In that program, measurements of the 
Ca II H and K lines of several thousand stars are made as a proxy for surface 
magnetism.

The observed quantity, {\it S}, is the flux measured in two 0.1 nm pass bands 
centered on the H and K lines normalized by two 2.0 nm-wide sections of 
photospheric flux centered at 390.1\,nm and 400.1\,nm.  A nightly calibration 
factor is determined from measurements of a standard lamp and standard stars 
(\cite{Baliunasetal95}).  The night-to-night rms precision of the lamp is on 
the order of 0.5\%, while the standard stars have an average standard 
deviation of $\sim 1.5\%$, which limits the lowest amplitude of variability 
that can be detected to approximately 1\%. 

Our periodogram analysis of the 141 Ca II observations over a period range of 
two to ten days gives a strong periodicity at $3.85\pm 0.01$\,d, essentially 
identical to the radial-velocity and photometric periods.  Since the time 
span of the Ca II observations is only 62 days and the period uncertainty is 
larger than for the radial velocities, we adopted the radial-velocity period 
and $T_0=2450996.5$ for the analysis. 

A 30-day subset of the {\it S} measurements, during which we had simultaneous 
photometry and radial-velocity data, is plotted in the second from the top 
panels of Fig.~8.  A sinusoidal fit to these {\it S} values gives an 
amplitude of 0.023 and a mean value of 0.46.  Interestingly, the phase 
is shifted 1/8 of a cycle compared to the light and color curves (with 15\% 
statistical error).  The variation of {\it S} is therefore 1/8 cycle out of 
phase with the radial-velocity variation.  The star has its higher level of 
Ca II H and K emission 1/8 of a cycle after it is faintest and 1/8 of a 
cycle before it reaches its most red-shifted radial velocity.  Similar 
phase differences between photometric and photospheric data are also detected
in the G0 Hyades star VB\,31 (\cite{Radick87}).  
 
\section{Discussion and modeling}

Our observations of HD~166435 indicate that the star is active and thus 
young.  The relatively fast rotation also indicates a young
star. If we assume that the X-ray flux detected by ROSAT (\cite{Burleigh97}) 
indeed originates from the star, the 0.5\,c/s PSPC counts in the 0.1-2.4keV 
range (\cite{Randich95}) corresponds to $\log(L_X)=29.5$ or $\lxlb=-4.1$.  
About the same X-ray luminosity is measured in other single stars of 
similar temperature in the 220\,Myr cluster NGC\,6475 (\cite{James97}). 
Thus, we estimage the age of HD\,166435 to be about 200\,Myr.

The typical Li content of G0 stars in the Pleiades is about 
$\log(N_{Li})\approx3.0$.  At the age of the Hyades (400\,Myr), the lithium 
is depleted to about 2.6.  The solar value is 1.0. The weak lithium content 
of HD~166435 noted above ($\log(N_{Li})<1.7$) conflicts with its 
young age deduced from its rotation rate and activity level.  We have no 
explanation for this.  However, other young stars with 
low lithium content have been detected in Preasepe (KW\,392) (\cite{Sod93b}) 
and Hyades (vB\,9, vB\,143) (\cite{Thorburn93}). 

The brightness of HD~166435, its spectroscopic line profiles, and its 
chromospheric emission all vary with the 3.8-day periodicity found in the
radial velocities.  As seen in Fig.\,8 above, the brightness variations
exhibit a roughly one-quater phase shift with respect to the radial 
velocities.  All of this strongly suggests that the radial-velocity variations
are due to the observed stellar activity.  For a magnetically active star 
with dark spots being carried into and out of view by rotation, we would expect 
to see such a phase shift between the brightness changes and radial-velocity
variations.  When the spot lies on the central meridian of the star, it
lies partly on the approaching and partly on the receeding halves of the
stellar disk.  The spot also causes the maximum depression in the brightness
of the star at this time.  Hence the radial-velocity effect of the spot 
at photometric minimum should be null.  As can be seen in Fig.\,8 above, 
the radial-velocity shift is very close to zero at the minimum of the 
light curve.  Radial-velocity minimum (the time of maximum blue shift) 
should occur as the dark spot approaches the receeding limb of the star
and hence is moving out of view.  This results in more light from the 
approaching half of the stellar disk reaching our telescopes than from the 
spotted, receeding half of the disk.  This is in agreement with Fig.\,8,
where we see that radial-velocity minimum corresponds to a time when the
star is rapidly brightening.  Similar reasoning predicts that radial-velocity
maximum should occur when the star is rapidly dimming, as also observed
in Fig.\,8.  The delta $(b-y)$ color curve in the bottom panel of the figure
shows that the star is redder when it is fainter, further showing that 
the spots causing the brightness variation must be cooler (darker) than
the surrounding photosphere.

The small phase shift between the brightness of HD~166435 and its 
chromospheric emission level suggests that the photocenter of the dark spot 
distribution is somewhat offset from the photocenter of the bright plage 
area.  This phenomena is observed on the Sun to a lesser degree where leader 
spots tend to be larger and to live longer than follower spots, resulting in 
a displacement up to 5$^0$ between the photocenters of the spot and the 
associated plages (\cite{DobsonHochey86}).  The larger displacement in 
HD~166435 suggests that the size of active regions is much larger than 
those seen on the Sun, in agreement with the observed amplitude of the 
light curve.

The growth and decay of individual spots on time scales longer than a
stellar rotation will affect the phase and amplitude of the radial-velocity 
signal (as well as the brightness and chromospheric emission levels) on 
long time scales.  Our detection of coherent radial-velocity variations 
without noticeable phase shifts over intervals of up to 21\,days suggests 
that the spots are stable for at least this long.  This provides an estimate 
of the evolutionary time scale for individual spots of roughly a month.
However, we observed a quasi-coherent signal for the entire two-year span 
of our radial-velocity measurements.  This suggests that new spot formation 
occurs at about the same longitude on the surface of the star during this
time.  Among the numerous active stars measured in our extra-solar planet 
survey, the stability of the radial-velocity signal of HD~166435 is unique.
None of the other active stars in our sample display the level of phase
coherence observed in HD~166435 that could suggest the possibility of
planetary reflex motion.  Therefore, the spot generation mechanism of 
HD~166435 may be the result of unusual stability in the geometry of its
magnetic field.  
This  phenomenon is not unique amongst active stars. 
\cite{Toner88} observed a stable phase lasting for at least 3 years in 
the variation of the line bissector   of 
the  young star $\chi$\,Boo\,A. 
Stable active regions at the same longitude are observed
as well in  RS CVs binaries (\cite{Jetsu96}).  Moreover, on the Sun, during flares, 
hot spots   have been seen for 30 years at the same longitude (\cite{Bai88}; \cite{Bai90})

We conclude from these qualitative arguments that a rotating spot model 
can successfully explain the variability in our spectroscopic and 
photometric data sets.  Moreover, the mean level of the Ca index, $<S>=0.46$,
corresponding to $R'_{HK}=-4.26$, is in a good agreement with the activity 
level expected for a star with rotation period of 3.8\,days (\cite{Noyes}). 

The combination of our rotation period and \vsini~ measurements allows us
to estimate that the inclination of the star's rotation axis  to our line of sight 
is approximately $i=30^0$. Notice that in our case the increase of 
the macroturbulence  with the
star rotation is negligible. It does not change  our estimate of  
the  $i$ angle.  The smooth (i.e., without plateaus) 
variations in our radial velocities, photometry, and chromospheric emission 
suggest that the largest spotted region is always visible to the observer 
and does not completely disappear when on the far side of the star.  
Therefore, it is likely that the spot region is located within 30 degrees 
of the stellar pole and is visible, at least in part, throughout the 
rotation cycle. In this case, projection effects would be responsible for
most of the rotational modulation of the star's brightness. 

Thus, we have shown that surface magnetic activity on the star HD~166435
can mimic the kind of radial-velocity variations observed in stars with
true planetary reflex motions.  However, further analysis of the
spectrcopic line profiles, photometry, and chromospheric emission clearly
demonstrates that stellar activity is the origin of the radial-velocity
variations.  Our observations of this peculiar object actually strengthen the 
interpretation of other low-mass companions to solar-type stars where a 
low-amplitude, radial-velocity variation has been detected but where 
photometric, chromospheric, and line-profile variations are absent
(e.g., \cite{henryetal00}).  

\begin{acknowledgements}

This work has been made possible thanks to the continuous support from the 
Fond National de la Recherche Suisse (FNRS) and the efforts made by the 
telescope staff of the Observatoire de Haute-Provence. We thanks Nuno Santos
and Claudio Melo for their contributions on  the  discussion 
of the  stellar activity.

We are grateful for the efforts of Lou Boyd and Don Epand at Fairborn 
Observatory.  Astronomy with automated telescopes at Tennessee State 
University is supported through NASA grants NCC5--96 and NCC5--511 and 
NSF grant HRD--9706268.  GWH acknowledges additional support from the 
Richard Lounsbery Foundation. 
 
The HK Project at Mount Wilson Observatory has been supported by the Richard 
Lounsbery Foundation, the Scholarly Studies Program and the James Arthur 
Funds of the Smithsonian Institution, The Olin Wilson Fund of the Mount Wilson 
Institute, MIT Space Grants \# 5700000633 and 7A06, NASA Grant NAG5--7635, 
and several generous individuals.  We are indebted to M. Bradford and K. 
Palmer for their dedication to the HK Project at Mount Wilson Observatory, 
which is operated under agreement with the Carnegie Institution of Washington.

\end{acknowledgements}


\newpage

\end{document}